\documentclass[preprint,showpacs,aps]{revtex4}

\usepackage{graphicx}
\usepackage{epsfig}
\usepackage{epstopdf}
\usepackage{dcolumn}
\usepackage{bm}
\usepackage{amsmath}
\usepackage{amssymb}
\usepackage{placeins} 

\def\Journal#1#2#3#4{{#1} {\bf#2}, {#3} {(#4)}}

\def\NP{{ Nucl. Phys.} }
\def\PLB{{ Phys. Lett.}  B}

\def\PRP{{ Phys. Rep.}}
\def\PRL{ Phys. Rev. Lett.}
\def\PR{{ Phys. Rev.}}
\def\PRD{{ Phys. Rev.} D}

\def\PRB{{ Phys. Rev.} B}
\def\PRA{{ Phys. Rev.} A}

\def\JPA{{J. Phys.} A}

\def\AP{Ann. Phys.} 
\def\RPP{Rep. Prog. Phys.}
\def\RMP{Rev. Mod. Phys.}

\def\vp{{\bf p}}

\def\be{\begin{equation}}
\def\ee{\end{equation}}
\def\bea{\begin{eqnarray}}
\def\eea{\end{eqnarray}}

\def\NP{{ Nucl. Phys.}}

\def\vp{\varphi}
\DeclareMathOperator{\PP}{PP}

\begin{document}
\title{The Casimir Effect in Spheroidal Geometries}

\author{A. R. Kitson}
\email{a.r.kitson@gmail.com}
\author{A. I. Signal}
\email{a.i.signal@massey.ac.nz}
\affiliation{Institute of Fundamental Sciences PN461 \\ Massey University \\
Private Bag 11 222,  Palmerston North 4442\\
New Zealand}

\vskip 0.5cm
\begin{abstract}
We study the zero point energy of massless scalar and vector fields subject to spheroidal boundary conditions.
For massless scalar fields and small ellipticity the zero-point energy can be found using both zeta function and Green's function methods. 
The result agrees with the conjecture that the zero point energy for a boundary remains constant under small deformations of the boundary that preserve volume (the boundary deformation conjecture), formulated in the case of an elliptic-cylindrical boundary.
In the case of massless vector fields, an exact solution is not possible. 
We show that a zonal approximation disagrees with the result of the boundary deformation conjecture.
Applying our results to the MIT bag model, we find that the zero point energy of the bag should stabilize the bag against deformations from a spherical shape.
\end{abstract}


\maketitle

\setcounter{footnote}{0}
\section{Introduction}

The Casimir effect is a manifestation of the quantum nature of fields. 
The presence of physical boundaries gives rise to changes in the zero point energy of a quantum field \cite{Cas48}. 
This has been studied theoretically for a number of geometries \cite{Milton01,KMM09}, however, apart from fairly simple cases such as parallel plates, spheres and cylinders, exact results are difficult to find because analytic solutions to the field equations are required.
The possibility of technological applications of quantum vacuum fluctuations \cite{BUM01,Lam05} has made further investigation of the Casimir effect in situations where the boundaries are either not perfectly conducting or of lower symmetry timely.

The problem of the Casimir self-stress on a spherical shell was first proposed by Casimir \cite{Cas56} as a possible mechanism for stabilizing a semiclassical model of the electron. 
However, Boyer showed \cite{Boy68} that the zero point energy is positive in this case. 
While the concept of the zero point energy of a single object must be carefully defined to avoid ambiguity, this notion has been applied to Kaluza-Klein theories \cite{AC83,KM87} and cosmological models, for example in \cite{Tag73,Eli01}.
Another area where the idea of the zero-point energy of a single object has found application is in hadronic physics. 
There have been many papers on the zero-point energy of coloured fields confined in a spherical cavity following the early work of Bender and Hays \cite{BH76} and Milton \cite{Mil80}.
Also, zero-point energy concepts have been applied to soliton \cite{FGJW00}, hybrid bag  \cite{CFKS00} and string models \cite{JKM04} of hadrons.

In recent years there has also been experimental and theoretical progress on the Casimir force between separated objects. 
We refer the reader to review articles by Lamoreaux \cite{Lam05}, Milton \cite{Mil09} and Klimchitskaya, Mohideen and Mostepanenko\cite{KMM092}.
Of particular note is the recent experimental finding \cite{MCP09} that the Casimir or, more properly, the Casimir - Lifshitz force can be repulsive, confirming a prediction of Lifshitz and collaborators \cite{DLP61} for a system where the two plates have different dielectric permittivities and the vacuum is replaced by a liquid with a permittivity whose value is between those of the plates.

It is well known that the Casimir energy is sensitive to geometry - even the sign of the energy is difficult to predict.
It is natural, therefore, to investigate the modification of zero point energy when the boundary is modified. 
Experimental investigations \cite{Chan08} have shown that changing the boundary on the nanometre scale alters the Casimir - Lifshitz force by a factor different from that predicted using the proximity force approximation, though more sophisticated calculations \cite{BE04,LM08} can account for the change in the force. 
Recent work \cite{CKMMM10} has provided experimental confirmation of an exact theoretical calculation of the effect of varying the amplitude of a sinusoidally corrugated surface.

Kitson and Romeo \cite{KR06} looked at the theoretical problem of an infinite cylinder deformed from circular to elliptical cross-section. 
For small ellipticity (or eccentricity) $e$ a perturbative-like expansion for the zero point energy can be found:
\bea
{\cal E}_{C}(a, e) = {\cal E}_{C}(a, 0) \left [ 1 + \frac{1}{2}e^{2} + {\cal O}(e^{4}) \right]
\eea
where $a$ is the semi-major axis of the ellipse, and ${\cal E}_{C}(a, 0)$ is the regularized zero point energy per unit length of an infinite circular cylinder of radius $a$, which has the numerical value \cite{DLM81}
\bea
{\cal E}_{C}(0) \approx -\frac{0.01356}{a^{2}}.
\eea
This agrees with earlier work of Kvitsinsky \cite{Kv96}.
Noting that an ellipse with semi-major axis $a$ and ellipticity $e$ has area $\pi a^{2} \sqrt{1-e^{2}}$, Kitson and Romeo's result can be written in terms of $R$, the geometric mean of the semi-major and semi-minor axes $R = \sqrt{ab}$
\bea
{\cal E}(e) \approx -\frac{0.01356}{R^{2}} \left [ 1 + {\cal O}(e^{4}) \right],
\eea
that is, if the cross-section of the cylinder remains constant under the deformation of the cross section from a circle of radius $R$ to an ellipse, then the zero point energy per unit length does not change, up to ${\cal O}(e^{4})$.
This is a remarkable result, and begs the question whether it can be generalized to other deformations of highly symmetric boundaries. 

In this paper we will consider the deformation of a sphere to a spheroid. 
It is known that the zero point energy for a spherical boundary is positive \cite{Boy68}, whereas it is negative in the cases of parallel plates and infinite circular and elliptical cylinder geometries. 
As these geometries are limiting cases of spheroidal geometry, it is interesting to investigate the change in zero point energy as a sphere is deformed into a spheroid, and to see whether this change lowers or increases the zero point energy. 
Some care may be needed here, as while the parallel plate configuration is the limiting case of an oblate spheroid with ellipticity taken to unity, there is also a change in topology, so the value of a physical quantity such as the zero point energy may not be continuous in this limit.

It should be recognized that the situation we are discussing is an idealization where boundary conditions have replaced a physical interaction, and care needs to be taken, especially with the appearance of divergences.
An alternative approach is to replace the boundary with a renormalizable coupling between the fluctuating field and a non-dynamical smooth background field \cite{BHR92,GJK03,REG09}.
This approach allows the use of the standard renormalization procedure of quantum field theory without boundaries to obtain finite results.

In section 2 we consider the case of a massless scalar field subject to Dirichelet and Neumann boundary conditions on the surface of a prolate spheroid. 
We obtain the zero point energy as a perturbative expansion in the ellipticity.
In section 3 we study the Maxwell equations with perfect conductor boundary conditions on the surface of a spheroid. 
Unfortunately an analytic solution for the fields is not possible in this case. 
In previous work \cite{KS06}, we proposed using a zonal approximation by weighting components of the known axial symmetric field solutions with a spheroidal boundary.
However, we will show that this approximation is poor, and likely to give the wrong sign for the first correction term in the perturbative expansion of the zero point energy.
In section 4 we examine the implications of our results for the MIT bag and flux tube models of hadron structure in QCD.

\section{Massless Scalar Field in Spheroidal Geometry}

The spheroidal boundary is given by the Cartesian equation
\bea
\frac{x^{2}}{a^{2}} + \frac{y^{2}+z^{2}}{b^{2}} = 1
\label{eq:sphbound}
\eea
where both $a$ and $b$ are non-zero.
A prolate spheroid has $a > b$ whereas an oblate spheroid has $a < b$.
We shall consider a prolate spheroid; the results for an oblate spheroid can be easily obtained from those we derive here.
We can think of the prolate spheroidal boundary as the solid figure formed by rotating the ellipse with semi-major axis $a$  and semi-minor axis $b$ about the $x$-axis.
The ellipticity of the spheroid is  
\bea
e = \sqrt{ \frac{a^{2}-b^{2}}{a^{2}}}
\eea

It is most convenient to work in prolate-spheroidal coordinates $(\xi, \eta, \phi)$ \cite{AS72}, which are related to Cartesian coordinates by
\bea
x&  = & f \xi \eta \nonumber  \\
y & = & f \sqrt{(\xi^{2}-1)(1-\eta^{2})} \cos \phi \nonumber \\
z & = & f \sqrt{(\xi^{2}-1)(1-\eta^{2})} \sin \phi
\label{eq:pspcoords}
\eea
where $f = \sqrt{a^{2}-b^{2}} = ea$.
The domains of the prolate spheroidal coordinates are $1 \leq \xi \leq \infty$, $-1 \leq \eta \leq 1$ and $0 \leq \phi < 2\pi$.
In these coordinates the boundary (\ref{eq:sphbound}) is given by
\bea
\xi = \frac{1}{e}.
\eea

Let us consider a real massless scalar field $\vp$ that satisfies the homogeneous Dirichelet condition on the boundary of the prolate spheroid.
In natural units the field equation and boundary condition are
\bea
(\nabla^{2} + \omega^{2}) \vp & = & 0 \\
\vp |_{\xi = 1/e} & = & 0 
\eea
where $\omega$ is the frequency.
In prolate spheroidal coordinates the field equation is separable, and the general expression for the field in the interior of the boundary can be found \cite{MF53}
\bea
\vp^{\rm I,D}(\xi, \eta, \phi) =  \sum_{l = 0}^{\infty} \sum_{m = - l}^{l} A_{l}^{m} S_{l}^{m (1)} (\xi , \gamma^{2}) 
{\rm ps}_{l}^{m (1)} (\eta , \gamma^{2}) \exp(i m \phi)
\label{eq:scalsoln}
\eea
where $\gamma = f \omega$.
Here the functions $S^{m(1)}_{l}$ are the radial prolate spheroidal functions of the first kind and ps$^{m(1)}_{l}$ are the angular prolate spheroidal functions of the first kind \cite{AS72,MS54}. 
The homogeneous Dirichelet boundary condition is satisfied when
\bea
S^{m(1)}_{l}(\frac{1}{e}, z^{2}e^{2}) = 0
\label{eq:spbc}
\eea
where we have introduced the dimensionless frequency $z = \omega a$.
Exterior to the boundary the general expression for the field is similar:
\bea
\vp^{\rm II,D}(\xi, \eta, \phi) = \sum_{l = 0}^{\infty} \sum_{m = - l}^{l} B_{l}^{m} S_{l}^{m (3)} (\xi , \gamma^{2}) 
{\rm ps}_{l}^{m (1)} (\eta , \gamma^{2}) \exp(i m \phi)
\eea
where $S^{m(3)}_{l}$ are the radial prolate spheroidal functions of the third kind. 

With these solutions to the field equation we can proceed to finding the zero point energy.
We will use two methods to find the zero point energy: the zeta function method and the Green's function method.

\subsection{Zeta Function Method}

The zeta function for the field is given by
\bea
\zeta(s) = \mu^{s} \sum_{\bf k} \omega_{\bf k}^{-s}
\eea
where $\omega_{\bf k}$ are the eigenenergies of the field, $s$ is a complex variable and $\mu$ is a positive mass scale.
The zero point energy is related to the principal part of the zeta function 
\bea
{\cal E}_{C} = \frac{1}{2} \PP_{s=-1} \mu \zeta(s)
\eea

Let the positive solutions of the spheroidal boundary condition equation (\ref{eq:spbc}) be given by $z_{lmn}$. 
For large enough Re$(s)$ we can write the zeta function for the interior field as 
\bea
\zeta^{\rm I,D}(s, a, e) = (\mu a)^{s} \sum_{l=0}^{\infty} \sum_{m=-l}^{l} \sum_{n=1}^{\infty} z_{lmn}^{-s}
\eea

The summation over $n$ can be performed via the argument principle \cite{WW62,TV05}
\bea
\sum_{n=1}^{\infty} z_{lmn}^{-s} = \frac{1}{2 \pi i} \int_{C} dz z^{-s} 
\frac{ S_{l}^{m(1) \prime}(1/e, z^{2}e^{2}) }{ S_{l}^{m(1)}(1/e, z^{2}e^{2}) }.
\eea
Here the prime denotes differentiation with respect to $z$, and the contour is a closed path that encloses all the positive zeroes of the radial spheroidal function and avoids the origin.

We can make progress via a formal asymptotic expansion of the radial prolate spheroidal functions for small ellipticity in terms of spherical Bessel functions of the first kind $j_{l}(z)$
\bea
S_{l}^{m(1)}(1/e, z^{2}e^{2}) & \sim & j_{l}(z) - \left[\frac{ l^{2} + l + m^{2} - 1 }{4 l^{2} + 4 l - 3} z j_{l}^{\prime} ( z )   \right.\nonumber \\
& & \qquad \qquad \left. - \left(\frac{ l^{2} + l - 3 m^{2} }{8 l^{2} + 8 l - 6} + \frac{m}{2} \right) j_{l} ( z )\right]e^{2} + 
{\cal O}(e^{4}). 
\label{eq:ferpsf}
\eea
The integrand of the zeta function can now be expressed using this expansion as 
\bea
\lefteqn{z^{- s} \frac{S_{l}^{m (1) \prime}(1 / e , z^{2} e^{2})}{S_{l}^{m (1)} (1 / e , z^{2} e^{2})} \sim  } \nonumber \\
& & z^{-s} \frac{j_{l}^{\prime} (z)}{j_{l}(z)} - \frac{l^{2} + l + m^{2} - 1}{4 l^{2}+ 4 l - 3} \left[ s z^{- s} \frac{j_{l}^{\prime} (z)}{j_{l} (z)} + \left( z^{1 - s} \frac{j_{l}^{\prime} (z)}{j_{l}(z)} \right)^{\prime} \right] e^{2} + O(e^{4}). 
\label{eq:zfexp}
\eea
The leading order term of (\ref{eq:zfexp}) can be recognized as the integrand we would obtain for a spherical boundary of radius $a$ \cite{LR96}. The ${\cal O}(e^{2})$ term has two terms: the first is proportional to the leading order term, the second is continuous on $C$ and has an antiderivative, so will integrate to zero. 
We can perform the summation over $m$ as the leading order term is independent of $m$, and for the next-to-leading order term we can use the relation
\bea
\sum_{m = - l}^{l} \frac{l^{2} + l + m^{2} - 1}{4 l^{2} + 4 l - 3} = \frac{2 l + 1}{3}.
\label{eq:msum}
\eea
So now we have that the next-to-leading order term is simply $-\frac{1}{3} s e^{2}$ times the leading order term, and we can write the zeta function as 
\bea
\zeta^{\rm I,D}(s, a, e) \sim \zeta^{\rm I,D}(s, a, 0)\left(1-\frac{s}{3}e^{2} + {\cal O}(e^{4})\right).
\label{eq:zeta1}
\eea

For the exterior field, we can do a similar analysis.
The formal asymptotic expansion of the radial prolate spheroidal functions $S^{m(3)}_{l}$ is the same as that for $S^{m(1)}_{l}$ except that the spherical Bessel functions of the first kind are replaced by spherical Hankel functions of the first kind $h^{(1)}_{l}$.
Thus we obtain the expansion of the zeta function in the exterior of the spheroidal boundary:
\bea
\zeta^{\rm II,D}(s, a, e) \sim \zeta^{\rm II,D}(s, a, 0)\left(1-\frac{s}{3}e^{2} + {\cal O}(e^{4})\right).
\eea

The total zeta function is the sum of internal and external contributions
\bea
\zeta^{\rm D}(s, a, e) & = & \zeta^{\rm I,D}(s, a, e) + \zeta^{\rm II,D}(s, a, e) \nonumber \\
& \sim & \zeta^{\rm D}(s, a, 0)\left(1-\frac{s}{3}e^{2} + {\cal O}(e^{4})\right).
\eea
The first term is the total zeta function for a massless scalar field satisfying the homogeneous Dirichelet condition on the boundary of a sphere of radius $a$. 
This term is well-behaved at $s = -1$, and taking the principal part gives the zero-point energy
\bea
{\cal E}^{\rm D}(a, e) \sim {\cal E}^{\rm D}(a, 0) \left( 1 + \frac{1}{3}e^{2} + {\cal O}(e^{4}) \right)
\label{eq:szpe}
\eea
where the leading term is the zero point energy of the massless scalar field for a sphere with Dirichelet boundary conditions \cite{BM94}
\bea
{\cal E}^{\rm D}(a, 0) = \frac{0.00281 \ldots}{a}
\eea

We can also consider Neumann (or Robin) boundary conditions.
The homogeneous Neumann boundary condition for the massless scalar field in prolate spheroidal geometry will be satisfied when the radial function satisfies
\bea
\frac{\partial}{\partial \xi}S^{m(1)}_{l}(\xi, u^{2}e^{2})|_{\xi = 1/e} = \tilde{S}^{m}_{l}(u, e) = 0,
\eea
where $u$ is now the dimensionless frequency.
We can find the zeta function for the interior region
\bea
\zeta^{\rm I,N}(s, e) = (\mu a)^{s}  \sum_{l=0}^{\infty} \sum_{m=-l}^{l} \sum_{n=1}^{\infty} u_{lmn}^{-s}.
\eea
As before, we can 
convert the summation over $n$ to a contour integral over complex $u$
\bea
\sum_{n=1}^{\infty} u_{lmn}^{-s} =\frac{1}{2 \pi i} \int_{C} du \, u^{-s} 
\frac{ \tilde{S}_{l}^{m \prime }(u, e) }{ \tilde{S}_{l}^{m}( u, e) }.
\eea
Here the prime denotes differentiation with respect to $u$.
Again using the formal expansion of the radial prolate spheroidal functions for small $e$, equation (\ref{eq:ferpsf}), we obtain an expression for the integrand of the zeta function which is very similar to equation (\ref{eq:zfexp}), except that the Bessel function $j_{l}$ is replaced by its derivative, the derivative $j_{l}^{\prime}$ is replaced by the second derivative, and we find an extra term which is a power of $u$:
\bea
\lefteqn{u^{- s} \frac{\tilde{S}_{l}^{m \prime}(u, e)}{\tilde{S}_{l}^{m} (u, e)} \sim  } \nonumber \\
& & u^{-s-1} + u^{-s} \frac{j_{l}^{\prime \prime} (u)}{j_{l}^{\prime}(u)} - \frac{l^{2} + l + m^{2} - 1}{4 l^{2}+ 4 l - 3} \left[ s u^{- s} \frac{j_{l}^{\prime \prime} (u)}{j_{l}^{\prime} (u)} + \left( u^{1 - s} \frac{j_{l}^{\prime \prime} (u)}{j_{l}^{\prime}(u)} \right)^{\prime} \right] e^{2} + O(e^{4}). 
\eea
As the contour avoids the origin, the $u^{-s-1}$ term will integrate to zero, and the analysis can proceed as before: the leading term being the same as for a spherical boundary, and the next-to leading-term having one part which integrates to zero and another part which is proportional to $-\frac{1}{3} s$ times the leading term. 
Hence we arrive at
\bea
\zeta^{\rm I,N}(s, a, e) \sim \zeta^{\rm I,N}(s, a, 0)\left(1-\frac{s}{3}e^{2} + {\cal O}(e^{4})\right).
\eea
Clearly this analysis extends to the field exterior to the prolate spheroid, and so we see that equation (\ref{eq:szpe}) holds for both Dirichelet and Neumann boundary conditions and can be considered a general result for the zero point energy of the massless scalar field.

\subsection{Green's function method}

The zero point energy is related to the reduced Green's function in a region of space 
\bea
{\cal E}_{C} = \frac{1}{2 \pi i} \int d\omega \int d^{3}x \, \omega^{2} g({\bf x, x}).
\label{eq:zpegf}
\eea

For the interior of the spheroidal boundary, with the homogeneous Dirichelet boundary condition, we have the reduced Green's function \cite{MF53}
\bea
\lefteqn{ {\rm g}^{\rm I,D} (\mathbf{x} , \mathbf{x'}) = } \nonumber \\
& & - i \omega \sum_{l = 0}^{\infty} \sum_{m = - l}^{l} X_{l}^{m} ( \eta , \phi , \gamma^{2} )
X_{l}^{m \ast} ( \eta ' , \phi ' , \gamma^{2} ) \frac{S_{l}^{m ( 3 )} (1 / e , \gamma^{2})}{S_{l}^{m ( 1 )} (1 / e , \gamma^{2})} 
S_{l}^{m ( 1 )} (\xi , \gamma^{2}) S_{l}^{m ( 1 )} (\xi ' , \gamma^{2}),
\eea
where
\bea
X_{l}^{m} ( \eta , \phi , \gamma^{2} ) = \sqrt{\frac{2 l + 1}{4 \pi} \frac{( l - m ) !}{( l + m ) !}} 
{\rm ps}_{l}^{m ( 1 )} (\eta , \gamma^{2}) \exp (i m \phi).
\eea
For the exterior region, the reduced Green's function is similar, except that the radial prolate spheroidal functions of the first and third kind are interchanged.
The spatial integral in equation (\ref{eq:zpegf}) becomes
\begin{equation}
\begin{split}
{\cal I}(z,e) & =  \int d^{3}x \, \omega^{2} {\rm g}^{\rm D}({\bf x, x})  \\
& =  - i \sum_{l = 0 }^{\infty} \sum_{m = - l}^{l} \int_{0}^{2 \pi} d \phi \int_{-1}^{1} d \eta X_{l}^{m} ( \eta , \phi , z^{2} e^{2} ) 
X_{l}^{m \ast} ( \eta , \phi , z^{2} e^{2} )  \\
& \quad \quad \times \left[ \frac{S_{l}^{m (3)} (1 / e , z^{2} e^{2})}{S_{l}^{m (1)} (1 / e , z^{2} e^{2})} \int_{z e}^{z} d \xi (\xi^{2} - \eta^{2} z^{2} e^{2}) \left( S_{l}^{m (1)} (\frac{\xi}{ze} , z^{2} e^{2}) \right)^{2} \right. \\
& \quad \quad \quad \left. + \frac{S_{l}^{m (1)} (1 / e , z^{2} e^{2})}{S_{l}^{m (3)} (1 / e , z^{2} e^{2})} \int_{z}^{\infty} d \xi (\xi^{2} - \eta^{2} z^{2} e^{2}) \left( S_{l}^{m (3)} (\frac{\xi}{ze} , z^{2} e^{2}) \right)^{2} \right],
\label{eq:spint}
\end{split}
\end{equation}
where we have rescaled $\xi$ by a factor of $ze$.
As in the previous subsection, we now make formal expansion of the radial and angular prolate spheroidal functions. 
Note that the argument of the spherical Bessel functions in the expansion (\ref{eq:ferpsf}) becomes $\xi$.
For  the angular prolate spheroidal functions we have the expansion for small $\gamma^{2}$
\bea
{\rm ps}_{l}^{m ( 1 )}(\eta , \gamma^{2}) &\sim P_{l}^{m}(\eta) + 
\left( \frac{( l + m - 1 ) ( l + m )}{2 ( 2 l - 1 )^{2} ( 2 l + 1 )} P_{l - 2}^{m} ( \eta  ) - 
\frac{( l - m + 1 ) ( l - m + 2 )}{2 ( 2 l + 1 ) ( 2 l + 3 )^{2}} P_{l + 2}^{m} ( \eta ) \right) \gamma^{2}  +  
{\cal O}( \gamma^{4} ).
\eea
We obtain the expansion of the spatial integral for small ellipticity
\bea
{\cal I}(z,e) \sim {\cal I}^{(0)}(z) + {\cal I}^{(2)}(z) e^{2} + {\cal O}(e^{4}).
\eea
As we expect, the leading term is the same as the spatial integral for the same problem on the boundary of a sphere of radius $a$ \cite{MDRS78}:
\begin{equation}
\begin{split}
{\cal I}^{(0)}(z) & = -i \sum_{l=0}^{\infty} \sum_{m=-l}^{l} \int_{0}^{2\pi} d\phi 
\int_{-1}^{1} d\nu \frac{2l+1}{4\pi}\frac{(l+m)!}{(l-m)!} \left( P_{l}^{m}(\nu) \right)^{2} \\
& \quad \quad \times
\left[\frac{h_{l}^{(1)}(z)}{j_{l}(z)} \int_{0}^{z} du \, u^{2} \left(j_{l}(u) \right)^{2} + 
\frac{j_{l}(z)}{h_{l}^{(1)}(z)} \int_{z}^{\infty} du \, u^{2} \left(h_{l}^{(1)}(u) \right)^{2}  \right]
\end{split}
\end{equation}
The next-to leading order term is given by
\bea
{\cal I}^{(2)}(z) = \lim_{e \rightarrow 0} \frac{1}{2} \frac{\partial^{2}}{\partial e^{2}} {\cal I}(z,e). 
\eea

We note that in the integral over the angular variables, only the leading term contributes because of the orthogonality of
the associated Legendre functions $P_{l}^{m}$. 
For the radial integrals, we need to use Leibnitz's rule. 
There is one term involving $\eta^2$, however this can be dealt with using the recurrence relation for associated Legendre 
functions
\bea
(2l+1)\eta P_{l}^{m}(\eta) = (l+1-m) P_{l+1}^{m}(\eta) +(l+m)P_{l-1}^{m}(\eta),
\eea
and the orthogonality relation.
After further simplification we can write
\bea
{\cal I}^{(2)}(z)  =  \frac{1}{3}{\cal I}^{(0)}(z) - \frac{1}{3} \frac{\partial}{\partial z} \left( z {\cal I}^{(0)}(z) + 
\sum_{l,m}{\cal Q}_{l}^{m}(z) \right).
\eea 
When the second term is integrated over $z$, we can use an $i\epsilon$ prescription to deform the contour in the complex plane to avoid the poles along the real axis, and close the contour around the upper half-plane. 
The integral will them vanish as it is a total derivative.
Thus the next-to-leading order term is $\frac{1}{3} e^{2}$ times the leading order term, which agrees with the above calculation using the zeta function method.

\subsection{Discussion}

Our equation (\ref{eq:szpe}) for the zero point energy of the prolate spheroid is a new result. 
We can generalize this result to an oblate spheroid. 
Rather than repeat our derivation above, we note that the ellipticity $e^{\prime}$ of an oblate spheroid is related to that of a prolate spheroid with $a \leftrightarrow b$ by
\bea
\frac{1}{e^{2}} + \frac{1}{e^{\prime 2}} =1.
\eea
Substituting into equation (\ref{eq:szpe}) yields
\bea
E^{\rm D}(a, e^{\prime}) \sim E^{\rm D}(a, 0) \left( 1 - \frac{1}{3}e^{\prime 2} + {\cal O}(e^{4}) \right),
\eea
similar to the result for a prolate spheroid.

We can gain more insight into our result by considering a sphere of radius $R$. 
The volume of this sphere is equal to that of the prolate spheroid when 
\bea
R = a \sqrt[3]{1 -  e^{2}},
\eea
which gives the zero point energy of the spheroid as 
\bea
{\cal E}(a, e) \sim {\cal E}(R, 0) \left( 1 +  {\cal O}(e^{4}) \right),
\eea
{\em i.\ e.\ } the leading correction term vanishes when the sphere is deformed to a spheroid in a manner that preserves the volume.
This agrees with the result of Kitson and Romeo for the elliptical cylinder geometry \cite{KR06}.
It is natural to speculate that this result may hold more generally \cite{Kit09}, though this is difficult to investigate. 
In the case of the elliptical cylinder, the result could be derived using a conformal map, which only requires continuity of the field solutions and is independent of boundary condition. 
For the spheroid we do not know the range of validity for our expansion in the ellipticity, so confirmation of this result may require experimental input. 

In the limit $e \rightarrow 1$, a prolate spheroid becomes a circular cylinder, which has negative zero point energy.
We note that our result for the zero point energy of the prolate spheroid is positive and increases in magnitude as the ellipticity increases. 
For an oblate spheroid, which becomes parallel plates in the limit $e^{\prime} \rightarrow 1$, our result for the zero point energy does decrease as the ellipticity increases, however, it remains positive. 
(A plot in an earlier paper \cite{KS06} erroneously showed a negative zero point energy.)
So the change in sign of the zero point as the geometry changes must come from higher order terms in the ellipticity expansion, or be associated with the change in topology at the limit.
In figure 1 we show the total zero-point energy for a scalar field subject to Dirichelet boundary conditions on both a prolate 
and an oblate spheroid.


\section{Electromagnetic Field in Spheroidal Geometry}

In general Maxwell's equations are not separable in spheroidal coordinates \cite{MF53}. 
This means that we cannot use either zeta function or Green's function methods to investigate the zero point energy of the electromagnetic field subject to appropriate boundary conditions on the boundary of either a prolate or oblate spheroid. 
However, there is one known case of the Maxwell equations separating in prolate spheroidal coordinates, which is when the field is axially symmetric \cite{Page49}.
In a previous paper we proposed that the zero point energy could be obtained in a zonal approximation by suitably weighting the zeta function of the axially symmetric solution, and obtained a result with the leading correction term in the ellipticity expansion having opposite sign to the spherical result \cite{KS06}. 
We now show that this result is unreliable, and that the zonal approximation is probably a poor conjecture.

In the case of a scalar field we can compare the exact result with the result of the zonal approximation. 
The zeta function for the field satisfying the homogeneous Dirichelet boundary condition on the prolate spheroid is given by \bea
\zeta^{\rm I,D}(s, a, e) = (\mu a)^{s}  \sum_{l=0}^{\infty} \sum_{m=-l}^{l} 
\frac{1}{2 \pi i} \int_{C} dz\, z^{-s} \frac{ S_{l}^{m(1) \prime}(1/e, z^{2}e^{2}) }{ S_{l}^{m(1)}(1/e, z^{2}e^{2}) }.
\label{eq:zetaID}
\eea
The zonal approximation is made by setting $m=0$ in the integrand and replacing the sum over $m$ by a factor $(2l+1)$.
We denote the zonal approximation to the zeta function by $\tilde{\zeta}$, and use the expansion in ellipticity, equation (\ref{eq:ferpsf}) to write 
\bea
\tilde{\zeta}(s, a, e) \sim \tilde{\zeta}^{(0)}(s, a) + \tilde{\zeta}^{(2)}(s, a) e^{2} + {\cal O}(e^{4}).
\eea
For simplicity we set the mass scale $\mu$ equal to $a$, and let $\nu = l + 1/2$. 
The leading order term is the zeta function for a spherical boundary
\bea
\tilde{\zeta}^{(0)}(s, a) = \sum_{l=0}^{\infty} \frac{2\nu}{2\pi i} \int_{C} dz\, z^{-s} \frac{j_{l}^{\prime}(z)}{j_{l}(z)},
\label{eq:zetasphere}
\eea
which has a Laurent expansion about $s=-1$ \cite{Rom95}
\bea
\tilde{\zeta}^{(0)}(s, a) = \frac{1}{315 \pi} \frac{1}{s+1} - 0.00889\ldots + {\cal O}(s+1).
\label{eq:zetarom}
\eea
The next-to-leading order term is found to be
\bea
\tilde{\zeta}^{(2)}(s, a) = -s \sum_{l=0}^{\infty} \frac{5-4\nu^{2}}{16(1-\nu^{2})}
\frac{2\nu}{2\pi i} \int_{C} dz\, z^{-s} \frac{j_{l}^{\prime}(z)}{j_{l}(z)}.
\label{eq:zetanlo}
\eea
This expression needs analytic continuation from the domain where it is absolutely convergent ${\rm Re}(s) > 2$ to 
$-1 < {\rm Re}(s) < 0$. 
This can be done by considering the contour to run along the imaginary axis from $z=+iy = +i\infty$ to $z=-iy=-i\infty$, avoiding the origin via a semi-circle of radius $\epsilon$, and being completed by a semi-circle in the right half-plane, see figure 2.
The line integrals along the large semi-circle, radius $\rho \rightarrow \infty$ and the small semi-circle, radius 
$\epsilon$ both vanish. 
The integral along the imaginary axis becomes
\bea
\tilde{\zeta}^{(2)}(s, a) = -s \sum_{l=0}^{\infty} \frac{5-4\nu^{2}}{8(1-\nu^{2})} \nu^{2-s} \frac{1}{\pi} \sin \left( \frac{\pi s}{2} \right)
\int_{0}^{\infty} dy \, y^{-s} \left( \frac{1}{\nu} \frac{i_{l}^{\prime}(\nu y)}{i_{l}(\nu y)} -1 + \frac{1}{2\nu y}\right)
\eea
where $i_{l}(z)$ are the modified spherical Bessel functions.
Now continuing to a neighborhood of $s = -1$ we find the Laurent expansion
\bea
\tilde{\zeta}^{(2)}(s, a) = \frac{3}{64 \pi} \frac{1}{(s+1)^{2}} - \frac{2561 -1890 \gamma - 5670 \ln 2}{40320 \pi} \frac{1}{s+1} 
-0.03421\ldots + {\cal O}(s+1)
\label{eq:leza}
\eea
where $\gamma = 0.57721 \ldots$ is the Euler-Mascheroni constant.
Taking the principal part of both the leading and next-to-leading terms, and factoring out the result for the spherical boundary, we have the zero point energy using the zonal approximation
\bea
\tilde{\cal E}^{\rm I,D}(a, e) \sim \tilde{\cal E}^{\rm I,D}(a, 0) \left( 1 -3.85312 \ldots e^{2} + {\cal O}(e^{4}) \right).
\eea

However, we can expand the zeta function, equation (\ref{eq:zetaID}), for small ellipticity as
\bea
\zeta^{\rm I,D}(s, a, e) \sim  \sum_{l=0}^{\infty} \sum_{m=-l}^{l}  \zeta^{\rm I,D}_{lm}(s, a, 0) 
\left( 1 - \frac{l^{2} + l + m^{2} - 1}{4 l^{2} + 4 l - 3} s e^{2} + {\cal O}(e^{4}) \right)
\eea
where $\zeta^{\rm I,D}_{lm}(s, a, 0)$ is the term appearing under the summation in equation (\ref{eq:zetasphere}) for the zeta function for the spherical boundary.
The summation over $m$ is performed using equation(\ref{eq:msum}) giving
\bea
\zeta^{\rm I,D}(s, a, e) \sim  \sum_{l=0}^{\infty} (2l+1) \zeta^{\rm I,D}_{lm}(s, a, 0) \left( 1 - \frac{s}{3}e^{2} +  {\cal O}(e^{4}) \right)
\eea
as expected from our earlier result, equation (\ref{eq:zeta1}).
Using the zeta function for the sphere, equation (\ref{eq:zetarom}) we get the exact result for the zero point energy 
\bea
{\cal E}^{\rm I,D}(a, e) \sim {\cal E}^{\rm I,D}(a, 0) \left( 1 + 0.25759 \ldots e^{2} + {\cal O}(e^{4}) \right).
\eea
We see that the zonal approximation gives a next-to-leading term that is the wrong sign and an order of magnitude too large.
Additionally, the Laurent expansion in the zonal approximation, equation (\ref{eq:leza}), has a double pole, which contradicts the Poisson kernel method.
We thus conclude that the zonal approximation is a poor approximation in this case.
We also note that if we had used the boundary deformation conjecture, equation (\ref{eq:szpe}), which takes into account the external field, the next-to-leading order term would be $+\frac{1}{3}{\cal E}^{\rm I,D}(a, 0)e^{2}$ which has the correct sign, and is not too different in magnitude from the exact result. 

In the coordinate systems where Maxwell's equations are separable ({\em e.\ g.\ } spherical polar or elliptical cylindrical coordinates), the field equations can be broken down into two sets of equations with solutions corresponding to {\em TE} or {\em TM} modes. 
For perfectly conducting boundary conditions, each of these modes (up to polarization states) is equivalent to a massless scalar field subject to either homogeneous Dirichelet or Neumann boundary conditions. 
So, in these geometries, finding the zero point energy of the massless scalar field with either Dirichelet or Neumann boundary conditions gives the zero point energy of the electromagnetic field as the sum of the zero point energies for the two solutions of the massless scalar field. 
In the case of the spheroid, we were able to find the zero point energy for both solutions of the massless scalar field, so it is interesting to conjecture that, for small ellipticity, the zero point energy of the electromagnetic field will follow our boundary deformation conjecture
\bea
{\cal E}^{\rm EM}(a, e) \sim {\cal E}^{\rm EM}(a, 0)  \left( 1 + \frac{1}{3}e^{2} + {\cal O}(e^{4}) \right)
\eea
where ${\cal E}^{\rm EM} (a, 0)$ is Boyer's result for a sphere of radius $a$
\bea
{\cal E}^{\rm EM} (a, 0) =\frac{0.04617 \ldots}{a},
\eea
which is the sum of zero point energies for Dirichelet and Neumann boundary conditions minus the contributions from $n=0$ modes.
The difficulty with proving this conjecture is that we do not know exactly what form the boundary conditions will take for the solutions of the field equations in either prolate or oblate spheroidal coordinates, as these solutions do not separate into `radial' and `angular' parts.
However, the simplicity of the perfectly conducting boundary conditions, which require only the `radial' part of the solution to vanish, or have vanishing first derivative, on the surface gives us some grounds for believing that these boundary conditions will not differ greatly from those we have used for the massless scalar field, and so the boundary deformation conjecture should be applicable for the electromagnetic field.

\section{Zero Point Energy in the MIT Bag Model}

The MIT bag model \cite{MIT74} is a phenomenologically successful model of hadron structure. 
In the model, the QCD  vacuum has infinite colour magnetic permeability $\mu$, while inside the bag the permeability is unity. 
This confines colour magnetic and electric fields to the interior with boundary conditions on the bag surface
\bea
\mathbf{n} \cdot \mathbf{E} = 0 = \mathbf{n} \times \mathbf{B}
\eea
where $\mathbf{n}$ is a unit vector pointing in the inward normal direction.
The original fit to hadron masses \cite{MIT75} included a term $-Z / R$ (where $r$ is the radius of a spherical bag) to account for zero point energy and centre of mass effects, with a value of $Z \approx 1.8$ giving a good fit.
We note that the fit for the strong coupling constant inside the bag typically gives fairly large values $\alpha_{S} \approx 2.2$ in order that the colour hyperfine splitting reproduce the nucleon-delta mass splitting of 300 MeV.

The zero-point energy of the confined colour fields, with divergences subtracted, can be estimated \cite{Mil80}
\bea
{\cal E}(R) \approx + \frac{0.7}{R}
\eea
which is of opposite sign to the value found in hadron mass fits. 
As the leading divergence is associated with the bag surface, more realistic boundary conditions incorporating a `skin depth', will not greatly alter this result \cite{Milton01}.
The contribution of zero point energy of fermion (quark and antiquark) fields is small, and is usually neglected.
Centre of mass motion is also estimated to give a positive contribution to hadron masses \cite{DJ80,Wong81}.
One possible resolution to this is to include further terms in the mass formula, such as a constant force term $FR$ \cite{Milton01,BVW88}, which allows reasonable fits to the hadron masses.

Another approach, in light of the large values of $\alpha_{S}$ found in the mass fits, is to consider non-perturbative effects. 
Recently Oxman, Svaiter and Amaral \cite{OSA05} considered modifying the gluonic Green's function to include confinement effects.
Based on Schwinger-Dyson analysis, they introduced a modified reduced Green's function
\bea
\check{g}(k^{2}) = \frac{(-k^{2})^{\lambda}}{(-k^{2}+\Lambda^{2})^{\lambda+1}}
\eea
where $\Lambda$ is a low momentum scale $\Lambda \sim \Lambda_{{\rm QCD}}$ and $\lambda$ is a positive constant, typically $\lambda \sim 1$.
This modified Green's function reproduces the expected large momentum behaviour $\check{g}(k^{2}) \rightarrow -1/k^{2}$, and has a power law behaviour at low momentum $\check{g}(k^{2}) \sim (-k^{2})^{\lambda}$.
The zero point energy can be expressed in terms of the reduced Green's function \cite{Milton01}
\bea
{\cal E} = -\frac{1}{2} \sum_{\mathbf k} \frac{1}{2\pi i} \int d\omega \, \ln  g(\omega^{2} - \omega_{\mathbf k}^{2}).
\eea
Now expressing the modified reduced Green's function in terms of the unmodified reduced Green's function
\bea
\check{g}(k^{2}) = g(k^{2})^{-\lambda} g(k^{2} - \Lambda^{2})^{\lambda+1}
\eea
we can find the modified zero point energy
\bea
\check{{\cal E}} = -\lambda {\cal E}(0) + (\lambda + 1){\cal E}(\Lambda^{2})
\eea
where ${\cal E}(m^2)$ is the zero point energy calculated using the unmodified Green's function for a field of mass $m$.
The second term will be small, so taking $\lambda \approx 1$ we obtain the modified result
\bea
\check{{\cal E}} \sim - \frac{0.7}{R}
\eea
which is in better agreement with the usual hadron mass fits.

Using the boundary deformation conjecture, we can investigate the stability of the spherical bag against deformations.
We note that the zero point energy of the bag no longer includes terms coming from the exterior of the boundary, so the boundary deformation conjecture is unlikely to be exact.
However, as we saw in the case of the massless scalar field in the interior region, the conjecture is able to give the correct sign and magnitude of next-to-leading terms in the ellipticity expansion.
Let us first consider a meson state (quark - antiquark pair). 
At small separations of the quark and antiquark, lattice QCD simulations \cite{Som86,BSS95,HSP96} show that the pair sit in a roughly spherical  potential, whereas at large separations the sphere becomes stretched into a flux tube of roughly the same radius as the original `bag'. 
We can model this as a spherical bag being deformed to a prolate spheroid with the semi-minor axis $b$ staying approximately constant.
Using the modified Green's function result above, we have the zero point energy of the spheroid
\bea
\check{{\cal E}} \sim -\frac{0.7}{a} \left( 1 + \frac{1}{3}e^{2} + {\cal O}(e^{4}) \right).
\eea
Now the semi-minor and semi-major axes of the spheroid are related by $b^{2} = a^{2}(1 - e^{2})$, so the zero point energy becomes
\bea
\check{{\cal E}}(e) \sim -\frac{0.7}{b} \left( 1 - \frac{1}{6}e^{2} + {\cal O}(e^{4}) \right),
\eea
so increasing the ellipticity will increase the energy of the bag.
We can also consider a baryon (three quark) state. 
At small inter-quark separations, the quarks sit in a spherical bag. 
Lattice QCD simulations \cite{BCKSLLW07} show that when one quark is separated from the other two a flux tube is formed, similar to the quark -antiquark case, with the flux tube again having approximately the same radius as the original bag. 
We can model this as a spherical bag being deformed to an oblate spheroid with constant semi-minor axis $a$.
The zero point energy of the oblate spheroidal bag using the modified Green's function result is
\bea
\check{{\cal E}} \sim -\frac{0.7}{a} \left( 1 - \frac{1}{3}e^{\prime 2} + {\cal O}(e^{\prime 4}) \right)
\eea
where $e^{\prime}$ is the ellipticity of the oblate spheroid.
Again we see that increasing the ellipticity increases the energy of the bag.
These results are consistent with the flux tube model result that the flux tube has constant energy per unit length.

\section{Summary}

In this paper we have investigated the zero point energy of a spheroid.
For a massless scalar field, subject to either Dirichelet or Neumann boundary conditions on the boundary of the spheroid, we were able to make an expansion in the ellipticity around $e = 0$ and find the lowest order correction term for deviations from a perfect sphere. 
We used both zeta function and Green's function methods to obtain the lowest order correction, and both methods agree. 
Our result is consistent with the boundary deformation conjecture of Kitson and Romeo \cite{KR06}, first seen in the case of an elliptical cylinder. 
In this case, the boundary deformation conjecture says that the zero point energy of a sphere does not change, up to 
${\cal O}(e^{4})$, when the sphere is deformed to a spheroid.
In the case of a massless vector field (or electromagnetism), we were able to show that the zonal approximation we had previously used to calculate the zero point energy \cite{KS06} is likely to be incorrect, and we argued that there are good reasons for believing that the boundary deformation conjecture should apply in this case, even if an exact solution is not possible. 
Finally, we applied the boundary deformation conjecture to the MIT bag model, where we argued that zero point energy  should increase when the bag is deformed from a sphere.

It is tempting to relate the boundary deformation conjecture to the results of Ambj\o rn and Wolfram \cite{AW83} for a cubiodal box boundary.
However, the results for the cuboid are only calculated for the field on the interior of the boundary, and are thus both divergent and dependent on the renomalization method used (zeta function regularization in this case), whereas our result for the spheroid (and the result of Kitson and Romeo for the elliptic cylinder geometry) includes both internal and external fields and is finite (as the divergences cancel) and unambiguous.
The finite part of Ambj\o rn and Wolfram's result shows that the sign of the zero point energy can change as the box is deformed, so it would be of considerable interest if their calculation could be extended to the external field, which would allow an unambiguous result.

It would also be of interest to extend the work of this paper to spinor fields and fields with mass. 
Finally, the ${\cal O}(e^{4})$ corrections in the ellipticity expansion should be calculated to enable us to place better limits on the boundary deformation conjecture.

\section*{Acknowledgments}

We would like to thank A. Romeo and K. A. Milton for useful discussions, and B. van Brunt for help with mathematical details. Financial support from Massey University and the Royal Society of New Zealand are gratefully acknowledged.

\section*{Appendix: Spheroidal Functions}

The scalar Helmholtz equation is separable in prolate spheroidal and oblate spheroidal coordinates.
As oblate spheroidal functions are easily related to prolate spheroidal functions \cite{AS72}, we shall only consider prolate spheroidal functions. 
The solution of the Helmholtz equation in prolate spheroidal coordinates has the form
\bea
\varphi(\xi, \eta, \phi) = \Xi(\xi) H(\eta) \exp (\pm i m \phi)
\eea
where the functions $\Xi$ and $H$ satisfy the ordinary differential equations
\bea
\left[ (\xi^{2} - 1) \frac{d^{2}}{d\xi^{2}} + 2\xi \frac{d}{d\xi} - \lambda_{lm}(\gamma^{2}) + \gamma^{2} (\xi^{2} - 1) - \frac{m^{2}}{\xi^{2} - 1} \right] \Xi(\xi) & = & 0  \label{eq:radspde} \\
\left[ (1 - \eta^{2}) \frac{d^{2}}{d\eta^{2}} - 2\eta \frac{d}{d\eta} + \lambda_{lm}(\gamma^{2}) + \gamma^{2} (1 - \eta^{2}) - \frac{m^{2}}{1 - \eta^{2}} \right] H(\eta) & = & 0. 
 \label{eq:angspde}
\eea
Here $\gamma = f \omega$, $m$ is an integer and $\lambda_{lm}$ is a constant of separation.
We note that these equations are identical, however, they apply over different ranges of the variables, $\xi \geq 1$ 
and $|\eta| \leq 1$.
The solutions to (\ref{eq:radspde}) 
are the radial prolate spheroidal functions; 
the solutions to (\ref{eq:angspde}) that are regular and well behaved at $\eta = \pm 1$ are the angular spheroidal functions of the first kind ${\rm ps}_{l}^{m(1)}(\eta, \gamma^{2})$ \cite{FAW03}.

\subsection*{Angular Prolate Spheroidal Functions}

Angular prolate spheroidal functions of the first kind can be expressed as a series of associated Legendre polynomials
\bea
{\rm ps}_{l}^{m(1)}(\eta, \gamma^{2})  =  \sum_{k=-\infty}^{\infty} (-1)^{k} a_{l,k}^{m}(\gamma^{2}) P_{l+2k}^{n}(\eta) 
\label{eq:apsexp} 
\eea
where the series coefficients satisfy a three term recurrence relation \cite{MS54,FAW03}
\bea
A_{l,k}^{m}(\gamma^{2}) a_{l,k-1}^{m}(\gamma^{2}) + \left(B_{l,k}^{m}(\gamma^{2}) - \lambda_{lm}(\gamma^{2}) \right) a_{l,k}^{m}(\gamma^{2}) + C_{l,k}^{m}(\gamma^{2}) a_{l,k+1}^{m}(\gamma^{2}) = 0
\eea
with coefficients
\bea
A_{l,k}^{m}(\gamma^{2}) & = & - \gamma^{2}\frac{(l-m+2k-1)(l-m+2k)}{(2m+4k-3)(2m+4k-1)} \nonumber \\
B_{l,k}^{m}(\gamma^{2}) & = & (l+2k)(l+2k+1) -2\gamma^{2}\frac{(l+2k)(l+2k+1)+m^{2}-1}{(2m+4k-1)(2m+4k-3)} \nonumber \\
C_{l,k}^{m}(\gamma^{2}) & = & - \gamma^{2}\frac{(l+m+2k+1)(l+m+2k+2)}{(2m+4k+3)(2m+4k+5)}.
\eea
There are a number of normalizations for the angular prolate spheroidal functions: we have used the normalization of reference \cite{MS54}
\bea
\int_{-1}^{1} d\eta \, {\rm ps}_{l}^{m(1)}(\eta, \gamma^{2}) {\rm ps}_{l^{\prime}}^{m(1)}(\eta, \gamma^{2}) = 
\frac{2}{2l+1} \frac{(l+m)!}{(l-m)!} \delta_{ll^{\prime}}.
\eea

In the spherical limit ($\gamma^{2} \rightarrow 0$), equation (\ref{eq:angspde}) becomes the associate Legendre equation with eigenvalue $\lambda_{lm} = l(l+1)$.
This enables us to make an expansion for the angular prolate spheroidal functions and their eigenvalues in powers of $\gamma^{2}$
\bea
{\rm ps}_{l}^{m(1)}(\eta, \gamma^{2})  & = & P_{l}^{m}(\eta) + \gamma^{2} Q_{lm}^{(2)} + {\cal O}(\gamma^{4}) \\
\lambda_{lm}(\gamma^{2})  & = & l(l+1) + \gamma^{2} \lambda_{lm}^{(2)} + {\cal O}(\gamma^{4}).
\eea
Substituting these expansions  back into the angular prolate spheroidal differential equation (\ref{eq:angspde}), and using the expansion in terms of associated Legendre polynomials (\ref{eq:apsexp}) it is straightforward, though tedious, to find 
\bea
{\rm ps}_{l}^{m ( 1 )}(\eta , \gamma^{2}) & \sim & P_{l}^{m}(\eta) + 
\left[ \frac{( l + m - 1 ) ( l + m )}{2 ( 2 l - 1 )^{2} ( 2 l + 1 )} P_{l - 2}^{m} ( \eta  )  \right.  \nonumber \\
& & \qquad \qquad \left. - \frac{( l - m + 1 ) ( l - m + 2 )}{2 ( 2 l + 1 ) ( 2 l + 3 )^{2}} P_{l + 2}^{m} ( \eta ) \right] \gamma^{2}  
 +  {\cal O}( \gamma^{4} ).
\label{eq:angapprox}
\eea
In figure 3 we plot the ratio of our expansion up to ${\cal O}(\gamma^{2})$ over the full angular prolate spheroidal function 
${\rm ps}_{l}^{0 ( 1 )}(\eta=0.5 , \gamma^{2})$ for $l = 0, 10, 20$ and 30 over the range $0 \leq \gamma^{2} \leq 1$. 
We see that the expansion is accurate to better than 1\% over this range of $\gamma$.

\subsection*{Radial Prolate Spheroidal Functions}

The solutions to the radial prolate spheroidal equation (\ref{eq:radspde}) are related to spherical Bessel functions.
If we take the limit $\gamma^{2} \rightarrow 0, \; \xi \rightarrow \infty$ with $z = \gamma \xi = $ constant, 
and also make the substitution $\Xi(\xi) = (1 - \xi^{-2})^{m/2} f(z)$, we find 
\bea
z^{2} \frac{d^{2}f}{dz^{2}} + 2z \frac{df}{dz} + (z^{2} - \lambda)f(z) = 0
\eea
which is satisfied by the spherical Bessel functions $j_{l}(z)$ and $y_{l}(z)$.
We are interested in the radial spheroidal functions of the first and third kinds, which satisfy the physical boundary conditions at the surface of the spheroid and at infinity.
These can be expressed as series of either spherical Bessel functions of the first kind $j_{l}(z)$ or spherical Bessel functions of the third kind $h_{l}(z) = j_{l}(z) + i y_{l}(z)$, also known as spherical Hankel functions of the first kind: \cite{FAW03}
\bea
S_{l}^{m(1,3)}(\xi, \gamma^{2})  =  \frac{(1 - 1/\xi^{2})^{m/2}}{A_{l}^{-m}(\gamma^{2})} \sum_{k=-\infty}^{\infty} 
a_{l, k}^{-m}(\gamma^{2}) f_{l+2k}(\gamma \xi) 
\label{eq:rpsexp} 
\eea
where $f = j, \; h$ respectively, and
\bea
A_{l}^{m}(\gamma^{2}) = \sum_{k=-\infty}^{\infty} (-1)^{k} a_{l, k}^{m}(\gamma^{2}).
\eea
The coefficients $a_{l, k}^{m}$ are the same as for the expansion of the angular prolate spheroidal functions.
The normalization has been chosen such that for large $\gamma \xi$
\bea
S_{l}^{m(1)}(\xi, \gamma^{2}) \rightarrow j_{l}(\gamma \xi).
\eea

Now we are able to make an asymptotic expansion in the ellipticity $e = 1/\xi$
\bea
S_{l}^{m(1)}(\frac{1}{e}, z^{2}e^{2})  \sim  j_{l}(z) + e^{2} T_{lm}^{(2)} + {\cal O}(e^{4}), 
\eea
and then by substituting back into the radial spheroidal equation (\ref{eq:radspde}) and using the Bessel function expansion (\ref{eq:rpsexp}) we can obtain 
\bea
S_{l}^{m(1)}(1/e, z^{2}e^{2}) & \sim & j_{l}(z) - \left[\frac{ l^{2} + l + m^{2} - 1 }{4 l^{2} + 4 l - 3} z j_{l}^{\prime} ( z )   \right.\nonumber \\
& & \qquad \qquad \left. - \left(\frac{ l^{2} + l - 3 m^{2} }{8 l^{2} + 8 l - 6} + \frac{m}{2} \right) j_{l} ( z )\right]e^{2} + 
{\cal O}(e^{4}). 
\label{eq:radapprox}
\eea
For the radial prolate spheroidal functions of the third kind $S_{l}^{m(3)}(1/e, z^{2}e^{2})$ we obtain the same expansion except that the spherical Bessel functions $j_{l}(z)$ are replaced by spherical Hankel functions of the first kind. 
In figure 4 we plot the ratio of our expansion up to ${\cal O}(e^{2})$ over the full radial prolate spheroidal function 
$S_{l}^{0(1)}(1/e, z^{2}e^{2})$ for $z = 10$ and $l = 0, 10, 20$ and 30 over the range $0 \leq e \leq 0.3$. 
We see that the expansion is accurate to better than 1\% for $e < 0.1$, but can become much less accurate as the ellipticity increases.

\FloatBarrier

\newpage
\begin{figure}[htb]
\begin{center}
\includegraphics[width=18 cm]{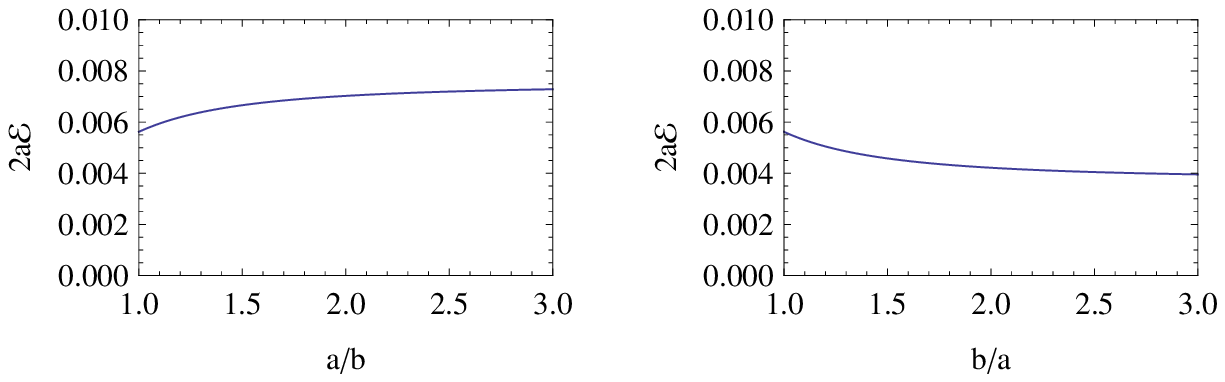}
\caption{Total zero-point energy of a scalar field with Dirichelet conditions on a prolate spheroid (left) and an 
oblate spheroid (right) of small ellipticity.}
\label{fig:scalarzpe}
\end{center}
\end{figure}
\newpage
\begin{figure}[htb]
\begin{center}
\caption[Integration contour]{The contour of integration for the zeta function in equation (\ref{eq:zetanlo}).}
\label{fig:c1c5}
\end{center}
\end{figure}

\newpage
\begin{figure}[htb]
\begin{center}
\includegraphics[width=10 cm]{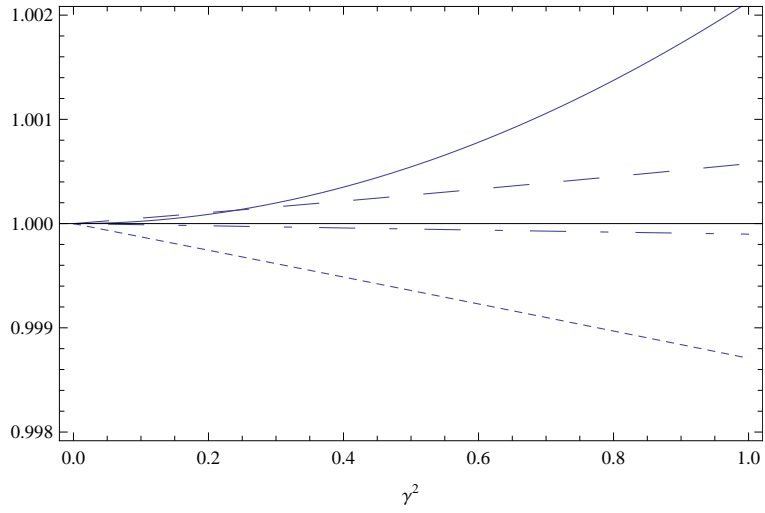}
\caption{The ratio of the approximation, equation (\ref{eq:angapprox}), to the full angular prolate spheroidal function ${\rm ps}_{l}^{0 ( 1 )}(\eta=0.5 , \gamma^{2})$ for $l = 0$ (full curve), 10 (dashed curve), 20 (dotted curve) and 30 (dot-dash curve).}
\label{fig:angapprox}
\end{center}
\end{figure}

\newpage
\begin{figure}[htb]
\begin{center}
\includegraphics[width=10 cm]{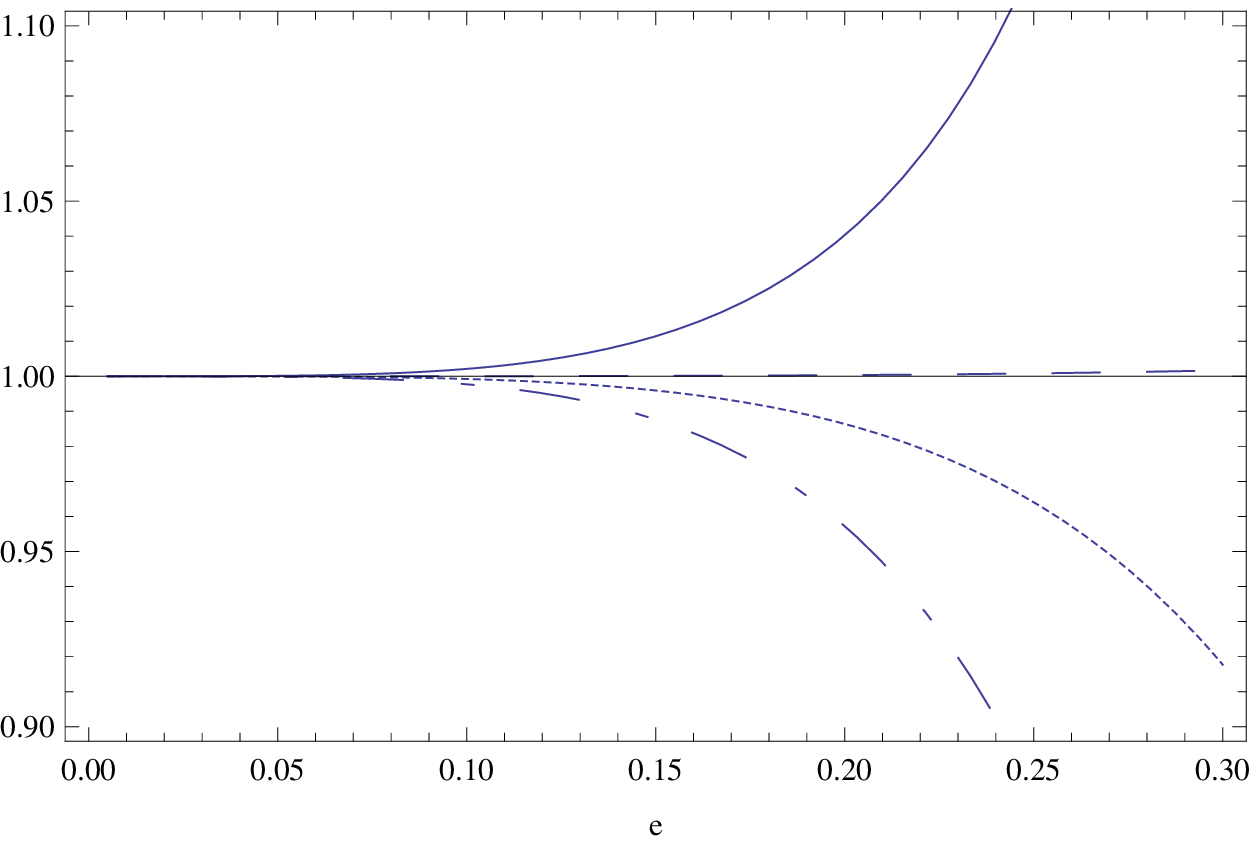}
\caption{The ratio of the approximation, equation (\ref{eq:radapprox}), to the full radial prolate spheroidal function $S_{l}^{0(1)}(1/e, z^{2}e^{2})$ for $z = 10$ and $l = 0$ (full curve), 10 (dashed curve), 20 (dotted curve) and 30 (dot-dash curve).}
\label{fig:radapprox}
\end{center}
\end{figure}


\begin{thebibliography}{99}

\bibitem{Cas48}
	H. B. G. Casimir,
	\Journal{Proc.\ Kon.\ Ned.\ Akad.\ Wetensch.}{51}{793}{1948}.

\bibitem{Milton01}
	 K. A. Milton,
	{\em The Casimir Effect}, World Scientific, Singapore (2001).

\bibitem{KMM09}
	 G. L. Klimchitskaya, U. Mohideen and V. M. Mostepanenko,
	{\em Advances in the Casimir Effect}, Oxford University Press, Oxford (2009).

\bibitem{BUM01}
	M. Bordag, U. Mohideen and V. M. Mostepanenko,
	\Journal{\PRP}{353}{1}{2001}.

\bibitem{Lam05}
	S. K. Lamoreaux,
	\Journal{\RPP}{68}{201}{2005}.

\bibitem{Cas56}	
	H. B. G. Casimir,
	\Journal{Physica}{19}{846}{1956}.
	
\bibitem{Boy68}
	 T. H. Boyer,
	\Journal{\PR}{174}{1764}{1968}.
	
\bibitem{AC83}
	 T. Appelquist and A. Chodos,
	\Journal{\PRL}{50}{141}{1983}.
	
\bibitem{KM87}
	 R. Kantowski and K. A. Milton,
	\Journal{\PRD}{35}{549}{1987}.
	
\bibitem{Tag73}
	 E. A. Tagirov,
	\Journal{\AP}{76}{561}{1973}.
	
\bibitem{Eli01}
	 E. Elizalde,
	\Journal{\PLB}{516}{143}{2001}.
	
\bibitem{BH76}
	 C. M. Bender and P. Hays,
	\Journal{\PRD}{14}{2622}{1976},

\bibitem{Mil80}
	 K. A. Milton,
	\Journal{\PRD}{22}{1441}{1980};
	\Journal{\PRD}{22}{1444}{1980}.
	
\bibitem{FGJW00}
	E. Fahri, N. Graham, R. L. Jaffe and H. Weigel,
	\Journal{\NP}{B585}{442}{2000};
	\Journal{\NP}{B630}{241}{2002}.
	
\bibitem{CFKS00}
	I. Cherednikov, S. Federov, M. Khalili and K. Sveshnikov,
	\Journal{\NP}{A676}{339}{2000}.

\bibitem{JKM04}
	K. J. Juge, J. Kuti and C. Morningstar,
	\Journal{\NP}{Proc. Supp. 129}{686}{2004};
	{\em Proc. Int. Conf. Color Confinement and Hadrons in QCD, Wako, Japan}, eds H. Suganuma, N. Ishii, M. Oka, 
	H. Enyo, T. Hatsuda, T. Kunihiro, K. Yazaki. World Scientific, Singapore (2004).

\bibitem{Mil09}
	K. A. Milton,
	\Journal{J.\ Phys.\ Conf.\ Ser.}{161}{012001}{2009}.

\bibitem{KMM092}
	 G. L. Klimchitskaya, U. Mohideen and V. M. Mostepanenko,
	\Journal{\RMP}{81}{1827}{2009}.
	
\bibitem{MCP09}
	J. N. Munday, F. Capasso annd V. A. Parsegian,
	\Journal{Nature}{457}{170}{2009}.
	
\bibitem{DLP61}
	I. E. Dzyaloshinskii, E. M. Lifshitz and L. P. Pitaevskii, 
	\Journal{Adv.\ Phys.\ }{10}{165}{1961}.
	
\bibitem{Chan08}
	H. B. Chan, Y. Bao, J. Zou, R. A. Cirelli, F. Klemens, W. M. Mansfield and C. S. Pai,
	\Journal{\PRL}{101}{030401}{2008}.

\bibitem{BE04}
	R. B\"uscher and T. Emig,
	\Journal{\PRA}{69}{062101}{2004};
	\Journal{\PRL}{94}{133901}{2005}.

\bibitem{LM08}
	A. Lambrecht and V. N. Marachevsky,
	\Journal{\PRL}{101}{160403}{2008}.

\bibitem{CKMMM10}
	H. C. Chiu, G. L. Klimchitskaya, V. N. Marachevsky, V. M. Mostepanenko and U. Mohideen,
	\Journal{\PRB}{81}{115417}{2010}.

\bibitem{BHR92}
	M. Bordag, D. Hennig and D. Robaschik,
	\Journal{\JPA}{25}{4483}{1992}.

\bibitem{GJK03}
	N. Graham, R. L. Jaffe, V. Khemani, M. Quandt, M. Scandurra and H. Weigel,
	\Journal{\NP}{B645}{49}{2002};
	\Journal{\PLB}{572}{196}{2003}.

\bibitem{REG09}
	S. J. Rahi, T. Emig, N. Graham, R. L. Jaffe, and M. Kardar,
	\Journal{\PRD}{80}{085021}{2009}.

\bibitem{KR06}
	A. R. Kitson and A. Romeo,
	\Journal{\PRD}{74}{085024}{2006}.

\bibitem{DLM81}
	J. DeRaad, L. Lester and K. A. Milton,
	\Journal{\AP}{136}{229}{1981}.
	
\bibitem{Kv96}	
	A. A. Kvitsinsky,
	\Journal{\JPA}{19}{649}{1996}.
	
\bibitem{KS06}
	A. R. Kitson and A. I . Signal,
	\Journal{\JPA}{39}{6473}{2006}.
	
\bibitem{AS72}
	 A. Abramowitz and I. A. Stegun,
	{\em Handbook of Mathematical Functions}, Dover, New York (1972).
	
\bibitem{MF53}
	P. M. Morse and H. Feshbach, 
	{\em Methods of Theoretical Physics}, McGraw-Hill, New York (1953).
	
\bibitem{MS54}
	J. Meixner and F. W. Sch\"{a}fka,
	{\em Mathieushe Funcktionen und Sph\"{a}roidfuncktionen}, Springer-Verlag,  (1954).
	
\bibitem{WW62}
	 E. T. Whittaker and G. N. Watson,
	{\em A Course of Modern Analysis}, Cambridge University Press, Cambridge (1962).

\bibitem{TV05}
	 W. Tutschke and H. L. Vasudeva,
	{\em An Introduction to Complex Analysis}, Chapman and Hall, Boca Raton (2005).
	
\bibitem{BM94}
	C. M.Bender and K. A. Milton,
	\Journal{\PRD}{50}{6547}{1994}. 	

\bibitem{LR96}
	S. Leseduarte and A. Romeo,
	\Journal{\AP}{250}{448}{1996}. 
	
\bibitem{MDRS78}
	K. A. Milton, L. L. DeRaad, Jr and J. S. Schwinger,
	\Journal{\AP}{115}{388}{1978}.	

\bibitem{Page49}
	L. Page,
	\Journal{\PR}{65}{98}{1944}.	
	
\bibitem{Kit09}
	A. R. Kitson, 
	{\em On the zero-point energy of elliptic cylindrical and spheroidal boundaries}, PhD thesis, Massey University (2009).

\bibitem{Rom95}
	A. Romeo,
	\Journal{\PRD}{52}{7308}{1995}.

\bibitem{MIT74}
	A. Chodos, R. L. Jaffe, K. Johnson, C. B. Thorn and V. Weisskopf,
	\Journal{\PRD}{9}{3471}{1974};
	A. Chodos, R. L. Jaffe, K. Johnson and C. B. Thorn ,
	\Journal{\PRD}{10}{2599}{1974}.

\bibitem{MIT75}	
	T. DeGrand, R. L. Jaffe, K. Johnson and J. Kiskis,
	\Journal{\PRD}{12}{2060}{1975}.

\bibitem{DJ80}	
	J. F. Donoghue and K. Johnson,
	\Journal{\PRD}{21}{1975}{1980}.

\bibitem{Wong81}	
	C. W. Wong,
	\Journal{\PRD}{24}{1416}{1981};
	K. F. Liu and C. W. Wong,
	\Journal{\PLB}{113}{1}{1982}.

\bibitem{BVW88}
	S. K. Blau, M. Visser and A. Wipf,
	\Journal{\NP}{B310}{163}{1988}.	

\bibitem{OSA05}	
	L. E. Oxman, N. F. Svaiter and R. L. P. G. Amaral,
	\Journal{\PRD}{72}{125007}{2005}.

\bibitem{Som86}	
	R. Sommer,
	\Journal{\NP}{B291}{673}{1986}.
	
\bibitem{BSS95}	
	G. S. Bali, C. Schlichter and K. Schilling,
	\Journal{\PRD}{51}{5165}{1995}.

\bibitem{HSP96}	
	R. W. Haymaker, V. Singh, Y. Peng and J. Wosiek,
	\Journal{\PRD}{53}{389}{1996}.
	
\bibitem{BCKSLLW07}	
	F. Bissey, F-G. Cao, A. R. Kitson, A. I. Signal, D. B. Leinweber, B. G. Lasscock and A. G. Williams,
	\Journal{\PRD}{76}{114512}{2007};
	F. Bissey, F-G. Cao, A. R. Kitson, B. G. Lasscock, D. B. Leinweber, A. I. Signal, A. G. Williams and J. M. Zanotti,
	\Journal{\NP}{B(Proc.\ Suppl.\ )141}{22}{2005}.
	
\bibitem{AW83}	
	J. Ambj\o rn and S. Wolfram,
	\Journal{\AP}{147}{1}{1983}.

\bibitem{FAW03}
	P. E. Fallon, P. C. Abbott and J. B. Wang,
	\Journal{\JPA}{36}{5477}{2003}.

\end{thebibliography}
\end{document}